\documentclass[12pt]{article}
\usepackage{epsf}
\def\beq{\begin{equation}}
\def\eeq{\end{equation}}

\begin{document}
\begin{titlepage}
\begin{center}
{\Large \bf William I. Fine Theoretical Physics Institute \\
University of Minnesota \\}  
\end{center}
\vspace{0.2in}
\begin{flushright}
TPI-MINN-06/03-T \\
UMN-TH-2428-06 \\
January 2006 \\
\end{flushright}
\vspace{0.3in}
\begin{center}
{\Large \bf Channel coupling in $e^+e^-$ annihilation into heavy meson pairs at the $D^* {\bar D}^*$ threshold
\\}
\vspace{0.2in}
{\bf M.B. Voloshin  \\ }
William I. Fine Theoretical Physics Institute, University of
Minnesota,\\ Minneapolis, MN 55455 \\
and \\
Institute of Theoretical and Experimental Physics, Moscow, 117259
\\[0.2in]
\end{center}

\begin{abstract}
A threshold singularity in a heavier meson channel, e.g. $D^* {\bar D}^*$, generally reflects in the amplitudes of $e^+e^-$ annihilation into lighter meson pairs, such as $D^* {\bar D}$, $D {\bar D}$, and $D_s {\bar D}_s$, due to the channel coupling. The behavior of the cross section of the $e^+e^-$ annihilation into each channel in this situation is considered using general properties of the amplitudes near the new threshold. It is argued that the effects of the channel coupling can be quite significant in the observable cross section, in particular near the $D^* {\bar D}^*$ threshold.
\end{abstract}

\end{titlepage}

It is quite well known\cite{pdg} that the cross section of $e^+e^-$ annihilation in the region just above the threshold of open charm production around 4.0 GeV displays an intricate behavior which is yet to be studied in detail. This behavior is contributed by the successive onset of specific channels with the $D$ mesons: $D {\bar D}$, $D^* {\bar D}$, $D_s {\bar D}_s$, etc., by the strong dynamics in each of these channels and by the coupling between them. Thus a detailed study of this region, as well as of the similar region at the $B^{(*)}$ meson threshold, may provide a wealth of information about the strong dynamics of systems with heavy and light quarks. It has been recognized long ago\cite{ov,drgg,nat,ek} that such dynamics can be quite interesting due to the fact that the strength and the range of the strong forces between the heavy mesons is fixed by the interaction of the light quarks and gluons, while the mass of the interacting constituents is set by the heavy quark mass. In particular this behavior suggests existence of bound or resonant states of ``molecular" type\cite{ov} at least in some meson-antimeson channels. A recent discovery\cite{belle1} of the resonance $X(3872)$ and its subsequent studies\cite{babar1,cdf,d0,belle2,babar2} most strongly support the interpretation\cite{cp,ps,mv,nat2} of this resonance as a ``molecular" $S$ wave state in the $D^0 {\bar D}^{*0}+ {\bar D^0}  D^{*0}$ channel with quantum numbers $J^{PC}=1^{++}$, which however is inaccessible in the $e^+e^-$ annihilation. 

It is quite apparent phenomenologically that the $P$ wave states of the meson pairs produced in the $e^+e^-$ annihilation have a nontrivial strong dynamics of their own. In particular both the $D {\bar D}$ and $B {\bar B}$ thresholds start with the all well known $\psi(3770)$ and $\Upsilon(4S)$ resonances, and there is a long known conspicuous enhancement of the $e^+e^-$ cross section near the $D^* {\bar D}^*$ threshold just above 4.00 GeV, which has been discussed long ago\cite{drgg} as a ``molecular" $D^* {\bar D}^*$ resonance. Thus existence of a resonance near the threshold of a specific heavy meson pair looks more like a rule than an exception, and it is quite possible that similar singularities will be observed at other thresholds, such as $D^* {\bar D}$,  $B^* {\bar B}$, etc., once a more detailed study of the $e^+e^-$ annihilation at those thresholds is done\cite{rosner}.

Admittedly, the current theoretical understanding does not provide much of an insight into either the properties, or even the very fact of existence of the threshold resonances. Furthermore, some of the usual guidance from the heavy quark limit is also lost in this situation. An example of such loss is the inapplicability of the conservation of the heavy quark spin at the energy scale relevant to the discussed resonances. Indeed, the part of the Hamiltonian, which depends on the chromomagnetic interaction of the heavy quark spin, is normally suppressed by the inverse of the heavy quark mass $m_Q$ and is of the order of $\Lambda_{QCD}^2/m_Q$, which is also the scale for the mass splitting between the vector and pseudoscalar heavy mesons. However at energies in the region of the thresholds of combinations of the vector and pseudoscalar mesons, the relevant overall energy scale is set by same mass splitting. Thus the spin-dependent Hamiltonian cannot be considered as a small perturbation in this region. This behavior in particular invalidates in this energy region any predictions of the relative yield of meson pairs with mesons of different spin, based on the heavy quark spin symmetry, as was demonstrated\cite{drgg} in the discussion of a threshold resonance in the  $D^* {\bar D}^*$ system within a simple spin correlation model, which model is equivalent in this channel to a more general scheme used in a coupled-channel model of charmonium\cite{cornell,elq}. It should be also mentioned that in the latter approach the breaking of the spin-counting rule and the structures in the cross section of the $e^+e^-$ annihilation are described\cite{cornell} in terms of the coupling between the charmed meson pairs and the charmonium dynamics rather than in terms of resonances in the charmed meson pairs, which approach is alternative to the one considered here.

In this situation it appears to be worthwhile to discuss the behavior of the production amplitudes across the specific meson-pair thresholds in a model independent way, using the general quantum mechanical properties of unitarity and analyticity. Although such approach inevitably involves phenomenological parameters, which currently can only be determined from data, this looks like a necessary step toward understanding the strong dynamics of the heavy hadrons. 
Furthermore, in view of a significant and rapid variation of the $e^+e^-$ cross section in the threshold region, a proper parametrization of the cross section is of a practical importance for unfolding the radiative corrections in a concrete detailed scan of this energy region, such as being undertaken in the CLEO-c experiment\cite{cleoc}. In this paper the behavior of the amplitudes for production of a lighter and a heavier meson pairs is considered immediately below and above the threshold of the heavier pair. For definiteness, the discussion will refer to the threshold of the `heavier' $D^* {\bar D}^*$ pair production, where the kinematically available `lighter' channels are $D {\bar D}$, $D^* {\bar D}$, and $D_s {\bar D}_s$, and where the situation is likely to be interesting due to apparent existence of a threshold resonance, although the discussed properties may also be relevant for other similar thresholds. It will be shown that a reflection, due to the channel coupling, of the $D^* {\bar D}^*$ threshold singularity in the lighter channels should give rise to a somewhat nontrivial behavior of the cross section in those channels. It should be mentioned that the reflection of the threshold onset in one channel on another coupled channel is well known (see e.g. in the textbook \cite{ll}, Sect. 147). However the specifics of the pure $P$ wave dynamics in the meson production amplitudes in the $e^+e^-$ annihilation and of the parameters pertinent to the $D^{(*)}$ mesons (and possibly to $B^{(*)}$ as well) bring in interesting features, which justify a separate study of this phenomenon in the discussed context.

In what follows the variation of the production amplitudes is considered in a rather narrow energy range of energy immediately below and above the $D^* {\bar D}^*$ threshold, on the order of about 30 MeV on each side. In this interval the $D^*$ mesons are nonrelativistic since their momentum $k$ (real above the threshold, and imaginary below the threshold) does not exceed approximately 250 MeV. Also such momenta are reasonably small in comparison with the inverse of the distance scale of the essential strong interaction, and the expansion of the dependence of the considered amplitude on the $D^*$ momentum can be limited to the first terms mandated by the general quantum-mechanical properties of the amplitudes. The lighter mesons, on the other hand, are already quite fast at the energy of the $D^* {\bar D}^*$ threshold, e.g. the momentum $p$ of each of the mesons in the $D^* {\bar D}$ pair is about 500 MeV, so that it cannot be considered as small. However the variation of this momentum over the considered narrow energy range is relatively weak, so that the proper dependence of the amplitudes on $p$ can be neglected in the first approximation, and we can rather concentrate on the dependence on the momentum $k$ whose variation is quite significant. 

There is a minor complication of the behavior in the threshold region due to the isotopic mass difference between the charged and the neutral $D^*$ mesons, which amounts to existence in fact of two separate thresholds split by about 7 MeV, and this splitting will be accounted for in the discussion here. Another potential complication arises from the Coulomb interaction between the charged $D^*$ mesons, described by the parameter $\alpha/v$, with $v$ being the c.m. velocity of the produced mesons. The effect of this interaction can be quite substantial at energies of about 5 MeV on each side of the threshold for the charged mesons. However, the Coulomb effects in the presence of strong interaction, especially in a situation, where there is a nearby resonance, have a peculiarity of their own\cite{mv3,mv4} and merit a separate study. In the present discussion these effects are ignored, so that the resulting formulas do not contain finer details, which may be present very close to the charged meson threshold.

We first start with considering the effects of the channel coupling ignoring the isotopic mass difference between the $D^*$ mesons, and will later modify the resulting formulas by including the mass splitting effects. It should be also mentioned that, although for the mesons produced in the $e^+e^-$ annihilation there is only one relevant partial wave, namely the $P$ wave\footnote{Generally an $F$ wave production amplitude is also possible for a pair of vector meson. However in the threshold region this higher partial wave can obviously be neglected.}, there are still several channels to be considered in the most general treatment. Namely, the `light' channels are $D^* {\bar D}$, $D {\bar D}$, and $D_s {\bar D}_s$, and for the opening at the considered energy channel $D^* {\bar D}^*$ there are still two possible states differing by the total spin of the vector meson pair: $S=0$ and $S=2$. The treatment of the coupling between the channels greatly simplifies in the presence of a resonance. In this case each of the production amplitudes $A_L$ in the light channels can be approximated as a (complex) constant $a_L$, the non-resonant background, and the contribution of the resonance, through which the coupling to the `heavy' channel takes place. As to the `heavy' channel itself, the non-resonant production is small near the threshold due to the $P$ wave kinematical suppression, and the amplitude can be approximated by that dominated by the resonance. The resonance couples to a specific linear combination of the two possible spin states, and the production of only this combination is relevant in the threshold region\footnote{Clearly, this reasoning rather naturally assumes that there is only one $J^{PC}=1^{--}$ resonance in or close to the considered energy region. The spin structure of the state of the $D^*$ mesons at the resonance can be determined experimentally from angular correlations\cite{ov2}.}, and the corresponding production amplitude for the $D^* {\bar D}^*$ pair is denoted here as $A_H$.

Let $W$ denote the total c.m. energy $E$ in the $e^+e^-$ annihilation relative to the threshold of production of the two vector mesons, each with the mass $M$: $W = E - 2M$, and let the `nominal' position of the resonance be at the complex energy corresponding to $W = W_0 - i \Gamma_0/2$. According to the general quantum-mechanical consideration, based on the unitarity and analyticity of the production and scattering amplitudes (see e.g. in the textbook \cite{ll}, Sects. 133 and 145) at energy above the `heavy' threshold, the resonant production amplitude $A_2$ has the form
\beq
A_H={b_H \, k \over W - W_0 + {i \over 2} \, ( \Gamma_0 + g_H^2 \, k^3)}~~~,
\label{ah}
\eeq
and the production amplitudes for each of the light channels are parametrized as
\beq
A_L=a_L + {b_L  \over W - W_0 + {i \over 2} \, ( \Gamma_0 + g_2^2 \, k^3)}~~~,
\label{al}
\eeq
where $k = \sqrt{M W}$ is the c.m. momentum of the $D^* {\bar D}^*$ pair. The coefficients $a_L, b_L,$ and $b_H$ are complex parameters, which are also taken to be constant in the present consideration. The magnitudes and the complex phases of these coefficients result from the rescattering among the `light' channels and the corresponding absorptive parts associated with these channels. It is worth noting that there is no reason to expect the phase of the coefficient $b_H$ to vanish at $k=0$, due to the existence of `light' inelastic channels strongly coupled to the `heavy' meson pair. The normalization convention used here for the amplitudes corresponds to the cross section  $\sigma (e^+e^- \to H) = |A_H|^2 \, k$ for the `heavy' channel and $\sigma (e^+e^- \to L) = |A_L|^2 \, p_L$ for each of the `light' channels with $p_L$ being the c.m. momentum in the corresponding `light' channel at the `heavy' threshold, which momentum is taken as a constant across the considered energy range.

The expressions (\ref{ah}) and (\ref{al}) can be readily found from the standard summation of the `blobs' in the graphs of the type shown in Fig.1, where the ultraviolet part of the loop is absorbed into the definition of the overall normalization and of the position $W_0$ of the resonance. However the general treatment of the threshold behavior of the amplitudes\cite{ll} guarantees that these expressions are quite independent of assumptions about the form factors used in the graphs of Fig.1.

\begin{figure}[ht]
  \begin{center}
    \leavevmode
    \epsfxsize=5in
    \epsfbox{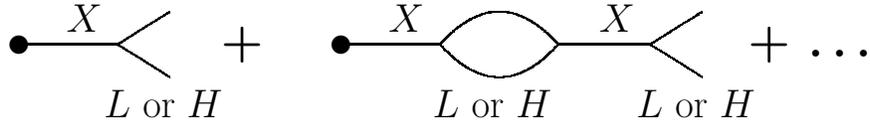}
    \caption{The graphs for the self energy of the resonance $X$ near the threshold for the `heavy' channel ($H$). The absorptive and dispersive parts of the self energy arise from loops with the `light' ($L$) and heavy meson pairs. The filled circle stands for the electromagnetic vertex.}  
  \end{center}
\end{figure}

Furthermore, the `nominal' width $\Gamma_0$ is due to the coupling to the lighter channels, and the term $g_H^2 \, k^3$ in the denominator in equations (\ref{ah}) and (\ref{al}) represents the additional contribution to the resonance width associated with the decay into the `heavy' meson pair, and $g$ is the coupling constant with the dimension of length. This extra contribution to the width can be written in terms of the excitation energy $W $ as $g^2 \, k^3 = W \, \sqrt{W/w}$, where 
\beq
w=g^{-4} \, M^{-3}
\label{wp}
\eeq 
is a parameter with dimension of energy. 
The critical point of the present consideration is that the amplitudes $A_L$ as functions of the energy $W$ are analytical functions of $W$ (with a cut at positive $W$) and the expression (\ref{al}) can be analytically continued to negative values of $W$. Using the notation $\Delta = - W$ at $W < 0$, the amplitudes $A_L$ below the `heavy' threshold take the form
\beq
A_L=a_L + {b_L  \over {1 \over 2} \Delta \sqrt{\Delta \over w} - \Delta - W_0 + {i \over 2} \,  \Gamma_0 }~~~.
\label{all}
\eeq

The term in the denominator in the latter formula, proportional to $\Delta^{3/2}$, is clearly the analytical continuation of the $g_H^2 \, k^3$ term in eq.(\ref{al}), and the proper sign is as indicated in eq.(\ref{all})\footnote{This sign is due to the $P$ wave motion, and is opposite to the one that would arise in a similar term for an $S$ wave ($\propto - \Delta^{1/2}$).}. This term becomes more essential than the one linear in $\Delta$, starting from $\Delta \approx w$, and gives rise to the effects discussed in the present paper. In particular, the expression ${1 \over 2} \Delta \sqrt{\Delta \over w} - \Delta$ cannot be arbitrarily negative and reaches its minimum, equal to $-16 \, w/27$ at $\Delta = \Delta_m= 16 \, w/9$. Thus in the case where the resonance is below the $D^* {\bar D}^*$ threshold, i.e. at negative $W_0=-\Delta_0$, the real part of the resonance denominator has either two zeros if $\Delta_0 < 16 w/27$, or no zero at all otherwise. In the latter case the position of the minimum of the absolute value of the denominator, i.e. of the maximal effect on the cross section, is at $\Delta=\Delta_m$ and does not depend on $\Delta_0$ at all, but rather is determined by the parameter $w$. The energy corresponding to $\Delta=\Delta_m$ in fact also determines the position of the maximum of the resonance amplitude in the case where the `nominal' position of the resonance is {\it above} the threshold, i.e. at positive $W_0$. This is due to the fact that at sufficiently small $w$ the damping of the `light' channel amplitude $A_L$ due to the absorptive term $g_H^2 \, k^3$ in the denominator in eq.(\ref{al}) suppresses the `light' channel above the `heavy' threshold.

Clearly, the significance of the discussed threshold effect critically depends on the value of the parameter $w$ for the near-threshold resonance. At least for two known resonances, $\psi(3770)$ and $\Upsilon(4S)$, at the thresholds of pairs of pseudoscalar heavy mesons the value of $w$ can be found from the data: $w \approx 60\,$ MeV for $\psi(3770)$, and $w \approx 20\,$ MeV for $\Upsilon(4S)$. There is every reason to expect that for the possible resonance near the $D^* {\bar D}^*$ threshold the coupling of the resonance to the vector meson pair should be stronger than for the case of pseudoscalar mesons, and the parameter $w$ should be significantly smaller than for $\psi(3770)$. Indeed, according to the data\cite{pdg}, the observed peak at the $D^* {\bar D}^*$ threshold\footnote{This peak is referred to in the Tables \cite{pdg} as $\psi(4040)$ with the maximum of the total $e^+e^-$ annihilation cross section at $4040\pm10$ MeV. However due to the threshold effect the maximum of the {\it total} cross section does not correspond to the position of the resonance, which position is in fact at a lower energy.} is mostly due to a very strong production of the $D^* {\bar D}^*$ pairs, and it has a width of $50 \pm 10$ MeV, in spite of a substantially smaller momentum $k$ of the vector mesons as compared to the momentum of the $D$ mesons at the peak of $\psi(3770)$. Thus the coupling $g$ for the vector mesons to their threshold resonance should be significantly stronger than of the pseudoscalar ones to $\psi(3770)$. The parameter $w$ is proportional to $g^{-4}$ (cf. eq.(\ref{wp})). Therefore, it is quite likely that for the $D^* {\bar D}^*$ threshold resonance this parameter can be smaller than that for the $\psi(3770)$ by a factor of 10 or more. In what follows it will be assumed that for the discussed resonance the value of $w$ is between 1 and 10 MeV.

As previously mentioned, for the actual $D^*$ mesons the isotopic mass difference gives rise to existence of two separate thresholds, the lower one for the neutral mesons and the upper one for the charged mesons. The energy difference\cite{pdg} $\delta=2 M(D^{*+})- 2 M(D^{*0})=6.60 \pm 0.14$ MeV between these thresholds is generally not very small on the discussed energy scale and it is appropriate to take it into account. For the illustrative estimates in this paper we assume that the coupling of the resonance to the $D^*$ mesons is isotopically symmetric and also ignore the Coulomb effects, so that the only difference between the channels with neutral and with charged mesons is purely kinematical. Using the previous notation $W$ for the excitation energy in the {\it neutral} channel: $W = E - 2 \, M(D^{*0})$, we write explicitly the production amplitudes in three separate regions of energy. Above both thresholds, i.e. at $W > \delta$:
\beq
A_{Hn}={1 \over \sqrt{2}} \, {b_H \, k_n \over W - W_0 + {i \over 4} \, \left [ 2\, \Gamma_0 + W \, \sqrt{W \over w} + (W-\delta) \, \sqrt{W-\delta \over w} \right ]}~~~,
\label{ahn3}
\eeq
\beq
A_{Hc}={1 \over \sqrt{2}} \, {b_H \, k_c \over W - W_0 + {i \over 4} \, \left [ 2\, \Gamma_0 + W \, \sqrt{W \over w} + (W-\delta) \, \sqrt{W-\delta \over w} \right ] }~~~,
\label{ahc3}
\eeq
\beq
A_L=a_L + {b_L  \over W - W_0 + {i \over 4} \, \left [ 2\, \Gamma_0 + W \, \sqrt{W \over w} + (W-\delta) \, \sqrt{W-\delta \over w} \right ] }~~~,
\label{al3}
\eeq
where the subscripts $n$ and $c$ refer to the channels with the neutral and the charged mesons. Between the thresholds, i.e. at $0 < W < \delta$, the production amplitudes take the form
\beq
A_{Hn}={1 \over \sqrt{2}} \, {b_H \, k_n \over {1 \over 4} (\delta-W) \, \sqrt{\delta -W \over w} + W - W_0 + {i \over 4} \, ( 2\, \Gamma_0 + W \, \sqrt{W \over w}  ) }~~~,
\label{ahn2}
\eeq
\beq
A_L=a_L + {b_L  \over  {1 \over 4} (\delta-W) \, \sqrt{\delta -W \over w} + W - W_0 + {i \over 4} \, ( 2\, \Gamma_0 + W \, \sqrt{W \over w}  ) }~~~.
\label{al2}
\eeq
Finally, below both thresholds, i.e. at $W < 0$, using the notation $\Delta =-W$ one can write
\beq
A_L=a_L + {b_L  \over  {1 \over 4} \left [ (\delta+\Delta) \, \sqrt{\delta + \Delta \over w}+ \Delta \, \sqrt{\Delta \over w} \right ]  - \Delta - W_0 + {i \over 2} \,  \Gamma_0  }~~~.
\label{al1}
\eeq

\begin{figure}[ht]
  \begin{center}
    \leavevmode
    \epsfxsize=6.5in
    \epsfbox{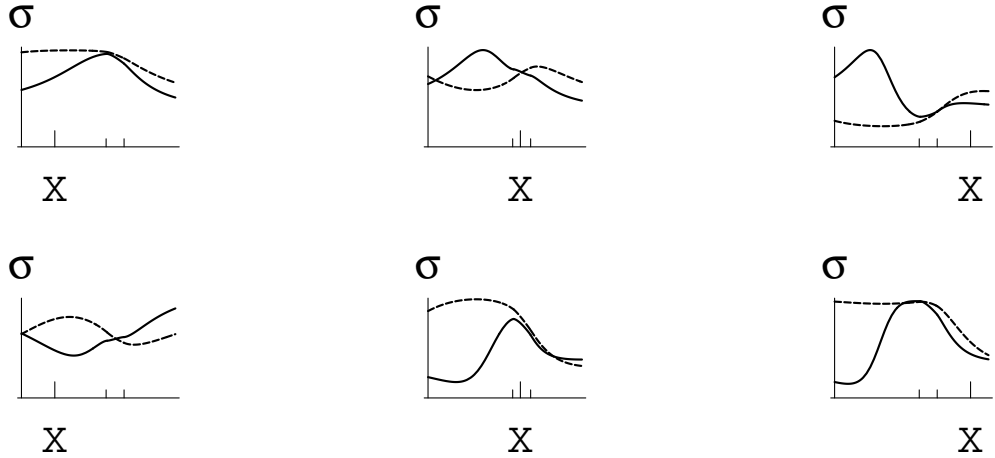}
    \caption{The types of behavior of the $e^+e^-$ annihilation cross section (in arbitrary units) into one of the `light' channels near the $D^* {\bar D}^*$ threshold. The horizontal axis in each plot spans the c.m. energy range from 3980 MeV to 4040 MeV. The $D^{*0} {\bar D}^{*0}$ and $D^{*+} {\bar D}^{*-}$ thresholds are shown with shorter vertical tick marks. The plots are shown for three assumed values of the `nominal' position $W_0$ of the resonance $X$ relative to the $D^{*0} {\bar D}^{*0}$ threshold: -20, 3, and 20 MeV, and the assumed value for this position is indicated in each plot by the longer vertical tick mark. In the plots in the upper row it is assumed that the coefficients $a_L$ and $b_L$ are relatively real and have the same sign. In the lower row of plots the relative sign of these coefficients is assumed to be negative. The solid lines in the plots correspond to $w=1\,$MeV, and the dashed one are for $w=10\,$MeV. The width parameter $\Gamma_0$ is fixed at $\Gamma_0=50\,$MeV. }  
  \end{center}
\end{figure}

That types of behavior of the cross section that can be observed in each of the `light' channels, i.e. $D^* {\bar D}$, $D {\bar D}$, or $D_s {\bar D}_s$, is illustrated for some representative values of the parameters. It is worth mentioning that the interference of the resonant amplitude with the non-resonant background is generally different for different channels, which may result in a significantly different behavior of the observed shape of the cross section for each channel. It is important to notice however that the features in the behavior of the cross section generally do not coincide with the position of the resonance. The shape of the cross section of the $e^+e^-$ annihilation into the vector  meson pairs $D^* {\bar D}^*$ above the threshold as found from the expressions (\ref{ahn3}), (\ref{ahc3}) and (\ref{ahn2}) is only weakly sensitive to the parameters of the resonance and is dominated by the very rapid rise from the threshold due to the $P$ wave factor $k^3$. 

Admittedly, at present the parameters involved in the discussed here description of the cross section are essentially unknown, and no prediction of specific shape of the cross section near the  $D^* {\bar D}^*$ threshold can be made.  However the presented here consideration illustrates that this energy region can be very feature-rich and a detailed scan of the $e^+e^-$ annihilation cross section into each of discussed channels with charmed meson pairs can provide a very interesting information on the parameters intimately related to the strong interaction of heavy mesons, an area for which the information is quite scant. Even the determination of the basic parameters of the possible resonance considered here should necessarily involve the channel coupling effects. For instance, the width parameter $\Gamma_0$, taken in the plots in Fig.2 at its PDG value\cite{pdg} of 50 MeV can in fact be quite different due to the significant distortion of the cross section due to the discussed effects.
An analysis of the angular distribution of the produced vector meson pairs would also determine the spin structure of the assumed threshold resonance. Finally, as mentioned previously, although the present discussion addresses mainly the  $D^* {\bar D}^*$ threshold, similar phenomena may take place also at other heavy meson pair thresholds, e.g. near the threshold region for  $D^* {\bar D}$ pairs at bout 3875 MeV, so that an exploratory test of those regions may also bring insight into new features of the heavy meson dynamics.

I thank Dan Cronin-Hennessy, Brian Lang and Ron Poling for enlightening discussions of the CLEO-c experiment. This work is supported in part by the DOE grant DE-FG02-94ER40823.

\end{document}